\documentclass[a4paper,11pt]{amsart}
\begin{document}

\hyphenation{gra-vi-ta-tio-nal re-la-ti-vi-ty Gaus-sian
re-fe-ren-ce re-la-ti-ve gra-vi-ta-tion Schwarz-schild
ac-cor-dingly gra-vi-ta-tio-nal-ly re-la-ti-vi-stic pro-du-cing
de-ri-va-ti-ve ge-ne-ral ex-pli-citly des-cri-bed ma-the-ma-ti-cal
de-si-gnan-do-si coe-ren-za pro-blem gra-vi-ta-ting geo-de-sic
per-ga-mon cos-mo-lo-gi-cal gra-vity cor-res-pon-ding
de-fi-ni-tion phy-si-ka-li-schen ma-the-ma-ti-sches ge-ra-de
Sze-keres con-si-de-red tra-vel-ling ma-ni-fold re-fe-ren-ces}

\title[On Hilbert's gravitational repulsion (\emph{A historical Note})]
{{\bf On Hilbert's gravitational repulsion \\ (\emph{A historical
Note})}}

\author[Angelo Loinger]{Angelo Loinger}
\address{A.L. -- Dipartimento di Fisica, Universit\`a di Milano, Via
Celoria, 16 - 20133 Milano (Italy)}
\author[Tiziana Marsico]{Tiziana Marsico}
\address{T.M. -- Liceo Classico ``G. Berchet'', Via della Commenda, 26 - 20122 Milano (Italy)}
\email{angelo.loinger@mi.infn.it} \email{martiz64@libero.it}

\vskip0.50cm

\begin{abstract}
In the literature on general relativity no mention is made of a
remarkable result contained in Hilbert's memoir \emph{Die
Grundlagen der Physik}, according to which in particular instances
and in particular regions the Einsteinian gravity exerts a
\emph{repulsive} action. We give here a concise illustration of
this peculiar phenomenon.
\end{abstract}

\maketitle



\normalsize

\vskip1.20cm \noindent \textbf{1.} -- Papers and treatises on
 general relativity do not mention an important and peculiar result
 by Hilbert (1917, 1924) \cite{1}, which concerns a \emph{repulsive}
 gravitational effect. We have made various applications of it
 \cite{2}. In the present Note, which has a historical character,
 we illustrate the essential of the Hilbertian argument, following
 faithfully the treatment of our Author (sect.\textbf{2}), and
 adding (sects. \textbf{3}, \textbf{4}) some remarks of a physical
 nature that intend to clarify the merits of the choice of
 coordinate time $t=x^{4}$ as evolution parameter for particular
 geodesics of Schwarzschild manifold.

\vskip1.20cm \noindent \textbf{2.} -- Hilbert's fundamental memoir
\emph{Die Grundlagen der Physik} \cite{1} ends with a detailed
treatment of various properties of Schwarzschild manifold created
by a point mass $m$ at rest \cite{3}. As it is well known
\cite{4}, the general expression of the relevant $\textrm{d}s^{2}$
is, if we employ spherical polar coordinates $r, \vartheta,
\varphi$:

\begin{eqnarray} \label{eq:one}
\textrm{d}s^{2} & = & \left[\frac{f(r)}{f(r)-2m}\right]
\left[\textrm{d}f(r)\right]^{2} + \left[f(r)\right]^{2}
\left(\textrm{d}\vartheta^{2}+\sin^{2}\vartheta \,
\textrm{d}\varphi^{2}\right) -
\nonumber\\
& & {} 
- \left[\frac{f(r)-2m}{f(r)}\right] \textrm{d}t^{2} \quad, \quad
(c=G=1) \quad,
\end{eqnarray}

where $f(r)$ is \emph{any} regular function of $r$ (of course,
under the condition that the $\textrm{d}s^{2}$ becomes
pseudo-Euclidean at spatial infinity). Schwarzschild's
\emph{original} $\textrm{d}s^{2}$ \cite{3} can be obtained by eq.
(\ref{eq:one}) putting $f(r)\equiv
\left[r^{3}+(2m)^{3}\right]^{1/3}$. Hilbert -- and, independently,
Droste \cite{5} and Weyl \cite{6} -- found a $\textrm{d}s^{2}$ for
which $f(r) \equiv r$, a form of solution that became the
\emph{standard form}, owing to its apparent simplicity.
Schwarzschild's field is \emph{maximally extended} and
diffeomorphic to the ``exterior part'' $r>2m$ of the standard
field. If we write, with Schwarzschild and Hilbert, $\alpha \equiv
2m$, the standard form is:

\begin{equation} \label{eq:oneprime}
\textrm{d}s^{2} = \frac{r}{r-\alpha} \, \textrm{d}r^{2} + r^{2}
\left(\textrm{d}\vartheta^{2}+\sin^{2}\vartheta \,
\textrm{d}\varphi^{2}\right) - \frac{r-\alpha}{r} \,
\textrm{d}t^{2} \quad. \tag{1\'{}}
\end{equation}

The geodesic lines of test-particles and light-rays in the field
$g_{jk}$, $(j, k=1, 2, 3,4)$, which characterizes the
$\textrm{d}s^{2}$ of eq. (\ref{eq:oneprime}), are exhaustively
investigated by Hilbert. Equations of motion and first integrals
are mainly, but not exclusively, written assuming an affine
parameter $p$, in lieu of the proper time $s$, as evolution
parameter; this allows a unified treatment of particles and
light-rays. For special purposes -- and \emph{pour cause} -- he
assigns to coordinate time $t=x^{4}$ the role of evolution
parameter.

\par The existence of  gravitational action of a \emph{repulsive}
nature reveals itself  in a very simple way for the
\emph{circular} orbits.  Hilbert's result -- which was
rediscovered by Einstein (1939, \cite{7}) in a different context
-- affirms that the value of the radial coordinate $r$  of a
circular trajectory of a particle must satisfy the following
inequality:

\setcounter{equation}{1}
\begin{equation} \label{eq:two}
r > \frac{3\alpha}{2}\quad;
\end{equation}

then, the passage from the affine parameter $p$ to coordinate time
$t=x^{4}$ as evolution parameter gives for the speed $v(t)$ a
formula which coincides with Newton's (remember that $c=G=1$):

\begin{equation} \label{eq:three}
v^{2} = \left(r \frac{\textrm{d}\varphi}{\textrm{d}t}\right)^{2} =
\frac{\alpha}{2r} \quad;
\end{equation}

by virtue of (\ref{eq:two}), we have:

\begin{equation} \label{eq:four}
v < \frac{1}{\sqrt{3}} \quad.
\end{equation}

For the circular geodesics of the light-rays, Hilbert found that
they satisfy necessarily the following equations:

\begin{equation} \label{eq:five}
r = \frac{3\alpha}{2}\quad,
\end{equation}

\begin{equation} \label{eq:six}
v = \frac{1}{\sqrt{3}}\quad.
\end{equation}

The above formulae are a clear expression of the gravitational
\emph{repulsion} exerted by our gravitating mass $m$.

\par Even more striking is the behaviour of the \emph{radial}
geodesics. The elimination of parameter $p$ from the Lagrangean
differential equation

\begin{equation} \label{eq:seven}
\frac{\textrm{d}}{\textrm{d}p} \left(\frac{2r}{r-\alpha} \,
\frac{\textrm{d}r}{\textrm{d}p}\right) +
\frac{\alpha}{(r-\alpha)^{2}} \, \left(
\frac{\textrm{d}r}{\textrm{d}p} \right)^{2} \, +
\frac{\alpha}{r^{2}} \, \left( \frac{\textrm{d}t}{\textrm{d}p}
\right)^{2} = \, 0 \quad,
\end{equation}

by means of the first integral

\begin{equation} \label{eq:eight}
\frac{r-\alpha}{r} \, \frac{\textrm{d}t}{\textrm{d}p} \, = \, 1
\quad,
\end{equation}

yields the following differential equation for $r$ as a function
of $t=x^{4}$:

\begin{equation} \label{eq:nine}
\frac{\textrm{d}^{2}r}{\textrm{d}t^{2}} -
\frac{3\alpha}{2\,r(r-\alpha)} \, \left(
\frac{\textrm{d}r}{\textrm{d}t} \right)^{2}
 \, + \, \frac{\alpha\,(r-\alpha)}{2\,r^{3}} = 0 \quad.
\end{equation}

Further, an analogous elimination of $p$ in the first integral

\begin{equation} \label{eq:ten}
\frac{r}{r-\alpha} \, \left( \frac{\textrm{d}t}{\textrm{d}p}
\right)^{2} - \frac{r-\alpha}{r} \, \left(
\frac{\textrm{d}t}{\textrm{d}p} \right)^{2} = A
\end{equation}

gives:

\begin{equation} \label{eq:eleven}
\left( \frac{\textrm{d}r}{\textrm{d}t} \right)^{2} =  \left(
\frac{r-\alpha}{r} \right)^{2} + A \,  \left( \frac{r-\alpha}{r}
\right)^{3} \quad,
\end{equation}

where $A$ is equal to zero for the light-rays and is a negative
constant for the material particles. Eq. (\ref{eq:nine}) tells us
that the acceleration is negative or positive -- \emph{i.e.}, the
gravitational force is attractive or repulsive -- in accord with
the following inequalities for the absolute value of the velocity:

\begin{equation} \label{eq:twelve}
\left|  \frac{\textrm{d}r}{\textrm{d}t} \right| <
\frac{1}{\sqrt{3}} \, \frac{r-\alpha}{r} \quad \textrm{--
(attraction)} \quad,
\end{equation}

or:

\begin{equation} \label{eq:thirteen}
\left|  \frac{\textrm{d}r}{\textrm{d}t} \right| >
\frac{1}{\sqrt{3}} \, \frac{r-\alpha}{r} \quad \textrm{--
(repulsion)}\quad,
\end{equation}

as it is not difficult to prove. For the light-rays, in
particular, we have from eq. (\ref{eq:ten}) with $A=0$ -- or from
$\textrm{d}s^{2}=0$ -- that

\begin{equation} \label{eq:fourteen}
\left|  \frac{\textrm{d}r}{\textrm{d}t} \right| =
\frac{r-\alpha}{r} \quad,
\end{equation}

\emph{i.e.}, by virtue of (\ref{eq:thirteen}), \emph{always
repulsion}; we see that light velocity increases from zero at
$r=\alpha$ to $1$ ar $r=\infty$.

\par If both $\textrm{d}r / \textrm{d}t$ and $\alpha$ are small, eq. (\ref{eq:nine})
gives the Newton equation

\begin{equation} \label{eq:fifteen}
\frac{\textrm{d}^{2}r}{\textrm{d}t^{2}} = - \frac{\alpha}{2} \,
\frac{1}{r^{2}} \quad.
\end{equation}

Of course, it is physically very important that eqs.
(\ref{eq:nine}) and (\ref{eq:eleven}) have as a consequence that
for $r=\alpha$ both velocity and acceleration are equal to zero;
\emph{no particle, no light-ray can overcome this ``barrier''}.

\vskip1.20cm \noindent \textbf{3.} -- We  have seen that the
existence of a repulsive gravitational action for the circular
geodesics is certified by inequality (\ref{eq:two}), which has
been derived through equations having an affine parameter as
evolution parameter. But for the radial geodesics Hilbert has
evidenced the repulsive action by means of equations having
$t=x^{4}$ as evolution parameter.

\par One could object that if we employ the proper time $s$ for
the geodesics of the particles, things stand otherwise. As a
matter of fact, the 4-velocity $\textrm{d}r / \textrm{d}s$ and the
4-acceleration $\textrm{d}^{2}r / \textrm{d}s^{2}$ of a
test-particle are, as it follows easily from Hilbert's formalism
by substituting $p$ with $s$:

\begin{equation} \label{eq:sixteen}
\left(\frac{\textrm{d}r}{\textrm{d}s} \right)^{2} =  E^{2} -
 \frac{r-\alpha}{r} \quad, \quad E=\textrm{a non zero constant} \quad,
\end{equation}

\begin{equation} \label{eq:seventeen}
\frac{\textrm{d}^{2}r}{\textrm{d}s^{2}} =  -
 \frac{1}{2} \, \frac{\alpha}{r^{2}} \quad \textrm{-- (attraction)} \quad,
\end{equation}

from which:

\begin{equation} \label{eq:sixteenprime}
\left(\frac{\textrm{d}r}{\textrm{d}s} \right)^{2}_{[r=\alpha]} =
E^{2} \quad, \tag{16\'{}}
\end{equation}

\begin{equation} \label{eq:seventeenprime}
\left(\frac{\textrm{d}^{2}r}{\textrm{d}s^{2^{}}}
\right)_{[r=\alpha]} = - \frac{1}{2\alpha} \quad, \tag{17\'{}}
\end{equation}

and it seems that the ``barrier'' $r=\alpha$ can be overcome.
Accordingly, it is necessary to make clear -- for the present case
-- the respective merits and defects of coordinate time $t=x^{4}$
and of proper time $s$.

 \vskip1.20cm \noindent \textbf{4.} -- First of all, we remember
 that in general relativity -- as it was explicitly emphasized by
 McVittie \cite{8} -- the interpretation of the four coordinates
 (mere ``labels'') of the point-events depends on the
 \emph{physical} situation contemplated by the investigation --
 exactly as it happens in the Lagrangean formulation of Newton's
 mechanics in regard to space coordinates.

 \par Now, the $g_{44}$ of eq. (\ref{eq:oneprime}) vanishes for
 $r=\alpha$. ``This means that a clock kept at this place would go at the rate
 zero.'' \cite{7}. This fact explains the results
 (\ref{eq:sixteenprime}) and (\ref{eq:seventeenprime}). Indeed, a
 clock aboard a test-particle goes gradually more and more slowly
 with its approaching to $r=\alpha$.

 \par On the contrary, the coordinate time $t=x^{4}$ of
 eq. (\ref{eq:oneprime}) is a \emph{global time} (``\emph{kosmische Zeit}''
 \cite{6}) of a static system: as it is well known \cite{9}, when
 $g_{0\alpha}=0$, $(\alpha=1, 2, 3)$, all the clocks in all the
 spatial points $(x^{1}, x^{2}, x^{3})$ can be synchronized.
 Accordingly, the rate of these clocks is physically more
 significant than the rate of the proper time of a test-particle,
 which suffers the influence of motion.

 \par We think that these considerations give a physical
 justification of Hilbert's use of $t=x^{4}$ as evolution
 parameter.

\vskip1.20cm \noindent \textbf{5.} -- A final remark. The original
Schwarzschild's form of solution \cite{3}, for which $f(r)\equiv
(r^{3}+\alpha^{3})^{1/3}$, $(\alpha\equiv 2m)$, or the very simple
Brillouin form \cite{10}, for which $f(r)\equiv r+\alpha$, are
\emph{maximally extended}. They are a proof of the essential
superfluity of Kruskal-Szekeres form of $\textrm{d}s^{2}$
\cite{11}. Further, this interval has various defects, in
particular:

\begin{itemize}
\item[\emph{i})] It entails a coordinate transformation from the
standard coordinates $(t, r, \vartheta, \varphi)$ to the new
coordinates $(v, u, \vartheta, \varphi)$, ``whose derivatives
happen to be singular at $r=\alpha$ in just the appropriate way
for providing a transformed metric that is regular there.''
\cite{12} --

\item[\emph{ii})] It gives a \emph{time-dependent} solution to a
\emph{static} problem --

\item[\emph{iii})] In a diagram $(u, v)$ the light-cones are
``\emph{open}'', as in \emph{special} relativity. This means the
abolition of any gravitational action on light-rays: \emph{a
consequence of i}) !
\end{itemize}

\par (Quite generally, the existence of singularities in a
coordinate transformation or/and in its derivatives can give
origin to a misrepresentation, or/and to a partial suppression of
the gravitational actions). --

\par In reality, Kruskal-Szekeres $\textrm{d}s^{2}$ has been
mainly utilized for a justification of the (meaningless) role
inversion of coordinate $r$ and coordinate $t$ in the ``interior
region'' $r<\alpha$ of Hilbert-Droste-Weyl potential $g_{jk}$ of
eq. (\ref{eq:oneprime}). Now, the mere existence of the Hilbertian
gravitational repulsion is sufficient to exclude any physical
meaning for this space region.

\vskip2.00cm
\begin{center}
\noindent \small \emph{\textbf{APPENDIX}}
\end{center}
\normalsize \noindent \vskip0.80cm

\begin{itemize}
\item[\textbf{\emph{$\alpha$}})] Hilbert's treatment of the
geodesic lines of test-particles and light-rays can be immediately
extended to the general form of $\textrm{d}s^{2}$ given by eq.
(\ref{eq:one}).

\par For instance, eq. (\ref{eq:two}) becomes $(\alpha\equiv 2m)$:

\begin{equation} \label{eq:A1}
f(r) > \frac{3}{2} \, \alpha \equiv 3m \quad; \tag{A1}
\end{equation}

thus, if $f(r)\equiv [r^{3}+(2m)^{3}]^{1/3}$ (see \cite{3}), we
obtain:

\begin{equation} \label{eq:A2}
r > 19^{1/3} \, m \approx 2.6684 \, m \quad; \tag{A2}
\end{equation}

and if $f(r)\equiv r+2m$ (see \cite{10}):

\begin{equation} \label{eq:A3}
r> 2m \quad; \tag{A3}
\end{equation}

for the isotropic coordinates we have $f(r)\equiv
\left(1+\frac{m}{2r} \right)^{2} r$, from which (see \cite{7}):

\begin{equation} \label{eq:A4}
r >  (2+\sqrt{3}) \, \frac{m}{2} \approx 1.86603  \,m \quad,
\tag{A4}
\end{equation}

if we employ with Fock \cite{13} harmonic coordinates, $f(r)\equiv
r+m$, from which:

\begin{equation} \label{eq:A5}
r > 2m \quad. \tag{A5}
\end{equation}

It is physically interesting that inequality (\ref{eq:A1}) is
stronger than inequality $f(r)>2m$.

\par Velocity $\textrm{d}[f(r)] / \textrm{d}t$ and acceleration
$\textrm{d}^{2}[f(r)] / \textrm{d}t^{2}$ of test-particles and
light-rays which travel along radial geodesics are equal to zero
when $f(r)=2m$. --

\item[\textbf{\emph{$\beta$})}] As it was emphazized by McVittie
many years ago \cite{14}, only solutions to Einstein field
equations concerning a \emph{single} mass at rest are known.
Therefore the Newtonian analogy is not sufficient for asserting
that a punctual mass could be a component of a binary system, or
that two punctual masses could collide. Indeed, an \emph{existence
theorem} would first be needed to show that Einstein equations
contain solutions which describe such configurations. Accordingly,
\emph{e.g.}, the interpretation of the observational data
concerning the quasar SDSSJ153636.22+044127 put forward by Boroson
and Lauer \cite{15} cannot be correct. --

\end{itemize}


\vskip1.80cm \small
\begin{thebibliography}{99}

\bibitem{1}
D. Hilbert, \emph{G\"ott Nachr.}, (1915) 395 (Erste Mitteilung,
vorgelegt am 20. Nov. 1915); \emph{Idem}, \emph{ibid.}, (1917) 53
(Zweite Mitteilung, vorgelegt am 23. Dez. 1916); \emph{Idem},
\emph{Math. Annalen}, \textbf{92} (1924) 1 -- also in
\emph{Gesammelte Abhandlungen}, Dritter Band (Verlag von J.
Springer, Berlin) 1935, p.258.

\bibitem{2}
A. Loinger and T. Marsico, \emph{arXiv:0706.3891 v3}
$[$physics.gen-ph$]$ 16 Jul 2007; \emph{ibid.}: \emph{0710.3927
v1} $[$\emph{id.}$]$ 21 Oct 2007; \emph{ibid.}: \emph{0711.4997
v3} $[$\emph{id.}$]$ 22 Dec 2007; \emph{ibid.}: \emph{0803.0050}
$[$\emph{id.}$]$ 1 Mar 2008; \emph{ibid.}: \emph{0809.122 v1}
$[$\emph{id.}$]$ 7 Sep 2008 (in this paper we show that also the
radial geodesics of Kerr manifold are subjected to Hilbert's
gravitational repulsion).

\bibitem{3}
K. Schwarzschild, \emph{Berl. Ber.}, (1916) 189; for an English
version see: \emph{arXiv:physics/9905030}, May 12th, 1999 -- and
\emph{Gen. Rel. Grav.}, \textbf{35} (2003) 951.

\bibitem{4}
A. S. Eddington, \emph{The Mathematical Theory of Relativity},
Second Edition (Cambridge University Press, Cambridge) 1960, p.94.
See also the Appendix in: L.S. Abrams, \emph{Phys. Rev.},
\textbf{20} (1979) 2474.

\bibitem{5}
J. Droste, \emph{Ned. Acad. Wet.}, Ser. \textbf{A}, \textbf{19}
(1917) 197.

\bibitem{6}
H. Weyl, \emph{Ann. Physik}, \textbf{54} (1917) 117; \emph{Idem},
\emph{Raum-Zeit-Materie}, Siebente Auflage (Springer-Verlag,
Berlin, \emph{etc.}) 1988, sect.\textbf{33}.

\bibitem{7}
A. Einstein, \emph{Ann. Math.}, \textbf{40} (1939) 922. In this
paper Einstein employs the isotropic coordinates whose
$\textrm{d}s^{2}$ can be obtained by putting $f(r)\equiv
\left(1+\frac{m}{2r} \right)^{2} r$ in eq. (\ref{eq:one}).

\bibitem{8}
G. C. Mc Vittie, \emph{General Relativity and Cosmology}
(University of Illinois Press, Urbana) 1965, sect.\textbf{4.1}.

\bibitem{9}
L. Landau et E. Lifchitz, \emph{Th\'eorie du Champ}, Deuxi\`eme
\'edition revue (\'Editions MIR, Moscou) 1966, sect.\textbf{84}.

\bibitem{10}
M. Brillouin, \emph{Journ. Phys. Rad.}, \textbf{23} (1923) 43; for
an English version see: \emph{arXiv:physics/0002009}, February
3rd, 2000.

\bibitem{11}
M. Kruskal, \emph{Phys. Rev.}, \textbf{119} (1960) 1743; G.
Szekeres, \emph{Publ. Mat. Debrecen}, \textbf{7} (1960) 285.

\bibitem{12}
S. Antoci and D.-E. Liebscher, \emph{Astr. Nachr.}, \textbf{322}
(2001) 137.

\bibitem{13}
V. Fock, \emph{The Theory of Space, Time and Gravitation}, 2nd
Revised Edition (Pergamon Press, Oxford, \emph{etc}.) 1964, p.215.

\bibitem{14}
G. C. Mc Vittie, \emph{The Observatory}, {\bf 98} (1972) 272.

\bibitem{15}
T.A. Boroson and T.R. Lauer, \emph{Nature}, \textbf{458} (2009)
53.

\end{thebibliography}
\end{document}